# A Comparison of Two Techniques for Bibliometric Mapping: Multidimensional Scaling and VOS

**Nees Jan van Eck and Ludo Waltman**
*Centre for Science and Technology Studies, Leiden University, The Netherlands and Econometric Institute, Erasmus School of Economics, Erasmus University Rotterdam, The Netherlands.*
*E-mail: {ecknjpvan, waltmanlr}@cwts.leidenuniv.nl.*

**Rommert Dekker**
*Econometric Institute, Erasmus School of Economics, Erasmus University Rotterdam, The Netherlands.*
*E-mail: rdekker@ese.eur.nl.*

**Jan van den Berg**
*Section of ICT, Faculty of Technology, Policy and Management, Delft University of Technology, The Netherlands.*
*E-mail: j.vandenberg@tudelft.nl.*

**VOS is a new mapping technique that can serve as an alternative to the well-known technique of multidimensional scaling. We present an extensive comparison between the use of multidimensional scaling and the use of VOS for constructing bibliometric maps. In our theoretical analysis, we show the mathematical relation between the two techniques. In our experimental analysis, we use the techniques for constructing maps of authors, journals, and keywords. Two commonly used approaches to bibliometric mapping, both based on multidimensional scaling, turn out to produce maps that suffer from artifacts. Maps constructed using VOS turn out not to have this problem. We conclude that in general maps constructed using VOS provide a more satisfactory representation of a data set than maps constructed using well-known multidimensional scaling approaches.**

## *Introduction*

In the field of bibliometrics and scientometrics, the idea of constructing science maps based on bibliographic data has intrigued researchers already for several decades. Many different types of maps have been studied. The various types of maps show relations among, for example, authors, documents, journals, or keywords, and they have usually been constructed based on citation, co-citation, or bibliographic coupling data or based on data on co-occurrences of keywords in documents. Quite some different techniques are available that can be used for constructing bibliometric maps. Without doubt, the most popular technique is the technique of multidimensional scaling (MDS).[1] MDS has been widely used for constructing maps of authors (e.g., McCain, 1990; White & Griffith, 1981; White & McCain, 1998), documents (e.g., Griffith, Small, Stonehill, & Dey, 1974; Small & Garfield, 1985; Small, Sweeney, & Greenlee, 1985), journals (e.g., McCain, 1991), and keywords (e.g., Peters & Van Raan, 1993a, 1993b; Tijssen & Van Raan, 1989). Recently, a new mapping technique was introduced that is intended as an alternative to MDS (Van Eck & Waltman, 2007a). This new mapping technique is called VOS, which stands for *visualization of similarities*. VOS has been used for constructing bibliometric maps in a number of studies (Van Eck & Waltman, 2007b, in press; Van Eck, Waltman, Noyons, & Buter, 2010; Van Eck, Waltman, Van den Berg, & Kaymak, 2006; Waaijer, Van Bochove, & Van Eck, 2010, in press).

An extensive comparison between the use of MDS and the use of VOS for constructing bibliometric maps does not yet exist. In this paper, we present such a comparison. We perform both a theoretical and an experimental analysis. In our theoretical analysis, we discuss the mathematics underlying MDS and VOS and we point out how the two techniques are mathematically related to each other. In our experimental analysis, we compare three approaches for constructing bibliometric maps. Two approaches rely on MDS, and the third approach relies on VOS. We use three data sets in our experimental analysis. One data set comprises co-citations of authors in the field of information science, another data set comprises co-citations of journals in the social sciences, and the third data set comprises co-occurrences of keywords in the field of operations research. Our experimental analysis indicates that maps constructed using either of the MDS approaches may suffer from certain artifacts. Maps constructed using the VOS approach do not have this problem. Based on this observation, we conclude that in general maps constructed using the VOS approach provide a more satisfactory representation of the underlying data set than maps constructed using either of the MDS approaches.

This paper is organized as follows. First, we discuss the use of MDS and VOS for constructing bibliometric maps and we study the mathematical relationship between the two techniques. Next, we present an experimental comparison of three approaches for constructing bibliometric maps, two approaches relying on MDS and one approach relying on VOS. Finally, we summarize the conclusions of our research.

## Multidimensional Scaling

In this section, we discuss the way in which MDS is typically used for constructing bibliometric maps. For more detailed discussions of MDS, we refer to Borg and Groenen (2005) and Cox and Cox (2001). From now on, we assume that the construction of bibliometric maps is done based on co-occurrence data (which includes co-citation data and bibliographic coupling data as special cases). We use the following mathematical notation. There are $n$ items to be mapped, which are denoted by 1, …, $n$. The items can be, for example, authors, documents, journals, or keywords. For $i \neq j$, the number of co-occurrences of items $i$ and $j$ is denoted by $c_{ij}$ (where $c_{ij} = c_{ji}$). The total number of co-occurrences of item $i$ is denoted by $c_i$. Hence, $c_i = \sum_{j \neq i} c_{ij}$ .

[1] Other techniques include the VxOrd technique (e.g., Boyack, Klavans, & Börner, 2005; Klavans & Boyack, 2006), the graph drawing techniques of Kamada and Kawai (1989) and Fruchterman and Reingold (1991), and the pathfinder network technique (e.g., Schvaneveldt, 1990; Schvaneveldt, Dearholt, & Durso, 1988; White, 2003). For overviews of various techniques, we refer to Börner, Chen, and Boyack (2003) and White and McCain (1997).

Below, we first discuss the calculation of similarities between items, and we then discuss the technique of MDS.

## Similarity Measures

MDS is usually not applied directly to co-occurrence frequencies. This is because in general co-occurrence frequencies do not properly reflect similarities between items (e.g., Waltman & Van Eck, 2007). To see this, suppose that journals A and B publish very similar articles. Suppose also that per year journal A publishes ten times as many articles as journal B. Other things being equal, one would expect journal A to receive about ten times as many citations as journal B and to have about ten times as many co-citations with other journals as journal B. It is clear that the fact that journal A has more co-citations with other journals than journal B does not indicate that journal A is more similar to other journals than journal B. It only indicates that journal A publishes more articles than journal B. Because of this, co-occurrence frequencies in general do not properly reflect similarities between items.

To determine similarities between items, co-occurrence frequencies are usually transformed using a similarity measure. Two types of similarity measures can be distinguished. Direct similarity measures (Van Eck & Waltman, 2009; also known as local similarity measures, see Ahlgren, Jarneving, & Rousseau, 2003) determine the similarity between two items by applying a normalization to the co-occurrence frequency of the items. Indirect similarity measures (also known as global similarity measures), on the other hand, determine the similarity between two items by comparing two vectors of co-occurrence frequencies. Most researchers interested in mapping authors or journals based on co-citation data rely on indirect similarity measures. Most other researchers rely on direct similarity measures. However, direct and indirect similarity measures can both be applied to any type of co-occurrence data. There is, for example, no reason to confine the use of indirect similarity measures to author and journal co-citation data.

Various direct similarity measures are being used in the literature. Especially the cosine and the Jaccard index are very popular. In a recent study (Van Eck & Waltman, 2009), we extensively analyzed a number of well-known direct similarity measures. We argued that the most appropriate measure for normalizing co-occurrence frequencies is the so-called association strength (e.g., Van Eck & Waltman, 2007b; Van Eck et al., 2006). This measure is also known as the proximity index (e.g., Peters & Van Raan, 1993a; Rip & Courtial, 1984) or as the probabilistic affinity index (e.g., Zitt, Bassecoulard, & Okubo, 2000). The association strength of items $i$ and $j$ is given by

$$\mathrm{AS}_{ij} = \frac{c_{ij}}{c_i c_j}. \tag{1}$$

It can be shown that the association strength of items $i$ and $j$ is proportional to the ratio between on the one hand the observed number of co-occurrences of $i$ and $j$ and on the other hand the expected number of co-occurrences of $i$ and $j$ under the assumption that co-occurrences of $i$ and $j$ are statistically independent (Van Eck & Waltman, 2009).

For a long time, the Pearson correlation has been the most popular indirect similarity measure in the literature (e.g., McCain, 1990, 1991; White & Griffith, 1981; White & McCain, 1998). Nowadays, however, it is well known that the Pearson correlation has some undesirable properties (Ahlgren et al., 2003; Van Eck &

Waltman, 2008). A well-known indirect similarity measure that does not have these undesirable properties is the cosine.[2] The cosine of items $i$ and $j$ is given by

$$\mathrm{COS}_{ij} = \frac{\sum_{k \neq i,j} c_{ik} c_{jk}}{\sqrt{\sum_{k \neq i,j} c_{ik}^2 \sum_{k \neq i,j} c_{jk}^2}}. \tag{2}$$

For a discussion of some other indirect similarity measures, we refer to an earlier paper (Van Eck & Waltman, 2008).

## The Technique of Multidimensional Scaling

After similarities between items have been calculated, a map is constructed by applying MDS to the similarities. The aim of MDS is to locate items in a low-dimensional space in such a way that the distance between any two items reflects the similarity or relatedness of the items as accurately as possible. The stronger the relation between two items, the smaller the distance between the items.

Let $s_{ij}$ denote the similarity between items $i$ and $j$ given by some direct or indirect similarity measure. For each pair of items $i$ and $j$, MDS requires as input a proximity $p_{ij}$ (i.e., a similarity or dissimilarity) and, optionally, a weight $w_{ij}$ ($w_{ij} \geq 0$). In the bibliometric mapping literature, the proximities $p_{ij}$ are typically set equal to the similarities $s_{ij}$. The weights $w_{ij}$ are typically not provided, in which case MDS uses $w_{ij} = 1$ for all $i$ and $j$. To determine the locations of items in a map, MDS minimizes a so-called stress function. The most commonly used stress function is given by

$$\sigma(\mathbf{x}_1, \ldots, \mathbf{x}_n) = \frac{\sum_{i<j} w_{ij} \left( f(p_{ij}) - \left\| \mathbf{x}_i - \mathbf{x}_j \right\| \right)^2}{\sum_{i<j} w_{ij} f(p_{ij})^2}, \tag{3}$$

where $f$ denotes a transformation function for the proximities $p_{ij}$ and $\mathbf{x}_i$ denotes the location of item $i$.[3] Typically, bibliometric maps have two dimensions and rely on the Euclidean distance measure. This means that $\mathbf{x}_i = (x_{i1}, x_{i2})$ and that

$$\left\| \mathbf{x}_i - \mathbf{x}_j \right\| = \sqrt{(x_{i1} - x_{j1})^2 + (x_{i2} - x_{j2})^2}. \tag{4}$$

As can be seen from Equation 3, MDS determines the locations of items in a map by minimizing the (weighted) sum of the squared differences between on the one hand the transformed proximities of items and on the other hand the distances between items in the map. For this idea to make sense, the transformation function $f$ has to be increasing when the proximities $p_{ij}$ are dissimilarities and decreasing when the proximities $p_{ij}$ are similarities.

Depending on the transformation function $f$, different types of MDS can be distinguished. The three most important types of MDS are ratio MDS, interval MDS, and ordinal MDS. Ratio and interval MDS are also referred to as metric MDS, while ordinal MDS is also referred to as non-metric MDS. Ratio MDS treats the proximities $p_{ij}$ as measurements on a ratio scale. Likewise, interval and ordinal MDS treat the proximities $p_{ij}$ as measurements on, respectively, an interval and an ordinal scale.[4] In ratio MDS, $f$ is a linear function without an intercept. In interval MDS, $f$ can be any linear function, and in ordinal MDS, $f$ can be any monotone function. We note that it makes no sense to use ratio MDS when the proximities $p_{ij}$ are similarities. This is because $f$ would then have to be a linearly decreasing function through the origin, which means that all transformed proximities would be negative or zero. In the bibliometric mapping literature, researchers often do not state which type of MDS they use. Since the proximities $p_{ij}$ are typically set equal to the similarities $s_{ij}$, ratio MDS does not seem to be used. Presumably, most researchers use ordinal MDS (e.g., McCain, 1990; White & Griffith, 1981; White & McCain, 1998).

---

[2] There are two different similarity measures, a direct and an indirect one, that are both referred to as the cosine. These two measures should not be confused with each other.

[3] The stress function in Equation 3 is referred to as the normalized raw stress function. Various alternative stress functions are discussed in the MDS literature (e.g., Borg & Groenen, 2005). In this paper, however, we do not consider these alternative stress functions. The normalized raw stress function is used by most MDS programs, including the PROXSCAL program in SPSS. Some MDS programs, such as the ALSCAL program in SPSS, use a somewhat different stress function.

The stress function in Equation 3 can be minimized using an iterative algorithm. Various different algorithms are available. A popular algorithm is the SMACOF algorithm (e.g., Borg & Groenen, 2005). This algorithm relies on a technique known as iterative majorization. The SMACOF algorithm is used by the PROXSCAL program in SPSS.

## VOS

In this section, we discuss the use of VOS for constructing bibliometric maps. The aim of VOS is the same as that of MDS. Hence, VOS aims to locate items in a low-dimensional space in such a way that the distance between any two items reflects the similarity or relatedness of the items as accurately as possible. As discussed below, VOS differs from MDS in the way in which it attempts to achieve this aim.

For each pair of items $i$ and $j$, VOS requires as input a similarity $s_{ij}$ ($s_{ij} \geq 0$). VOS treats the similarities $s_{ij}$ as measurements on a ratio scale. The similarities $s_{ij}$ are typically calculated using the association strength defined in Equation 1 (e.g., Van Eck & Waltman, 2007b; Van Eck et al., 2006). VOS determines the locations of items in a map by minimizing

$$V(\mathbf{x}_1, \ldots, \mathbf{x}_n) = \sum_{i<j} s_{ij} \left\| \mathbf{x}_i - \mathbf{x}_j \right\|^2 \tag{5}$$

subject to

$$\frac{2}{n(n-1)} \sum_{i<j} \left\| \mathbf{x}_i - \mathbf{x}_j \right\| = 1. \tag{6}$$

Hence, the idea of VOS is to minimize a weighted sum of the squared distances between all pairs of items. The squared distance between a pair of items is weighed by the similarity between the items. To avoid trivial solutions in which all items have the same location, the constraint is imposed that the average distance between two items must be equal to one.

There are two computer programs in which the VOS mapping technique has been implemented. Both programs are freely available. A simple open source program is available at [www.neesjanvaneck.nl/vos/](www.neesjanvaneck.nl/vos/), and a more advanced program called

[4] For a discussion of the concepts of ratio scale, interval scale, and ordinal scale, see Stevens (1946).

VOSviewer (Van Eck & Waltman, in press) is available at www.vosviewer.com. The two programs both use a variant of the SMACOF algorithm mentioned above to perform the minimization of Equation 5 subject to Equation 6.

We note that the objective function in Equation 5 has an interesting property.[5] To show this property, we introduce the idea of the ideal location of an item. We define the ideal location of item *i* as

$$\mathbf{x}_i^* = \frac{\sum_{j \neq i} s_{ij} \mathbf{x}_j}{\sum_{j \neq i} s_{ij}}. \quad (7)$$

That is, the ideal location of item *i* is defined as a weighted average of the locations of all other items, where the location of an item is weighed by the item's similarity with item *i*. (Notice the analogy with the concept of center of gravity in physics.) The ideal location of an item seems to be the most natural location an item can have. Because of this, it seems desirable that items are located as close as possible to their ideal location. This is exactly what the objective function in Equation 5 seeks to achieve. To see this, suppose that the locations of all items except item *i* are fixed, and ignore the constraint in Equation 6. Minimization of the objective function can then be performed analytically and results in $\mathbf{x}_i$ being equal to $\mathbf{x}_i^*$ defined in Equation 7. Hence, if the locations of all items except item *i* are fixed and if the constraint is ignored, minimization of the objective function causes item *i* to be located exactly at its ideal location. Of course, items do not have fixed locations, and solutions are determined not only by the objective function but also by the constraint. For these reasons, items will in general not be located exactly at their ideal location. However, due to the objective function, items at least tend to be located close to their ideal location.

## Relationship Between Multidimensional Scaling and VOS

In this section, we study the mathematical relationship between MDS and VOS. We show that, under certain conditions, MDS and VOS are closely related.

As discussed above, when researchers use MDS for constructing bibliometric maps, they usually seem to rely on ordinal or interval MDS. However, when MDS is applied to similarities calculated using the association strength defined in Equation 1, the use of ordinal or interval MDS is not completely satisfactory. This can be seen as follows. Suppose that items *i* and *j* have twice as many co-occurrences as items *i* and *k*. Suppose also that the total number of co-occurrences of item *j* equals the total number of co-occurrences of item *k*. Calculation of similarities using the association strength then yields $s_{ij} = 2s_{ik}$. Based on this, it seems natural to expect that in a map that perfectly represents the co-occurrences the distance between items *i* and *j* equals half the distance between items *i* and *k*. Of course, due to the inherent limitations of a low-dimensional Euclidean space, a map in which co-occurrences are perfectly represented usually cannot be constructed. However, ordinal and interval MDS do not even try to construct such a map. This is because in some sense the transformation function *f* has too much freedom in these types of MDS. In ordinal MDS, for example,

[5] Mapping techniques based on the objective function in Equation 5 have also been proposed by Belkin and Niyogi (2003) and by Davidson, Hendrickson, Johnson, Meyers, and Wylie (1998). However, the constraints used by these researchers are different from the constraint in Equation 6. In our experience, the constraint in Equation 6 yields much more satisfactory results than the alternative constraints used by other researchers.

$f$ can be any monotonically decreasing function, which means that any map in which the distance between items $i$ and $j$ is smaller than the distance between items $i$ and $k$ may serve as a perfect representation of the equality $s_{ij} = 2s_{ik}$. Hence, ordinal MDS may be indifferent between, for example, a map in which the distance between items $i$ and $j$ equals exactly half the distance between items $i$ and $k$ and a map in which the distance between items $i$ and $j$ is just slightly smaller than the distance between items $i$ and $k$.

We now propose an alternative way in which MDS can be applied to similarities calculated using the association strength (or to any other similarities that can be treated as measurements on a ratio scale). Our alternative approach does not have the above-mentioned disadvantage of ordinal and interval MDS. In our approach, we choose the transformation function $f$ to be simply the identity function, which means that $f(p_{ij}) = p_{ij}$. Using this transformation function, it is easy to see that minimization of the stress function in Equation 3 is equivalent with minimization of

$$\hat{\sigma}(\mathbf{x}_1,\ldots,\mathbf{x}_n) = \sum_{i<j} w_{ij} \left\| \mathbf{x}_i - \mathbf{x}_j \right\|^2 - 2\sum_{i<j} w_{ij} p_{ij} \left\| \mathbf{x}_i - \mathbf{x}_j \right\|. \tag{8}$$

Equation 8 makes sense only if the proximities $p_{ij}$ are dissimilarities. Because of this, we cannot set the proximities $p_{ij}$ equal to the similarities $s_{ij}$. Instead, we first have to convert the similarities $s_{ij}$ into dissimilarities $d_{ij}$. Converting similarities into dissimilarities can be done in many ways. We use the conversion given by $d_{ij} = 1 / s_{ij}$. This conversion has the natural property that if in a perfect map the distance between one pair of items is twice as large as the distance between another pair of items, the similarity between the first pair of items is twice as low as the similarity between the second pair of items. Substitution of $p_{ij} = d_{ij} = 1 / s_{ij}$ in Equation 8 yields

$$\hat{\sigma}(\mathbf{x}_1,\ldots,\mathbf{x}_n) = \sum_{i<j} w_{ij} \left\| \mathbf{x}_i - \mathbf{x}_j \right\|^2 - 2\sum_{i<j} w_{ij} \frac{1}{s_{ij}} \left\| \mathbf{x}_i - \mathbf{x}_j \right\|. \tag{9}$$

If two items $i$ and $j$ do not have any co-occurrences with each other, Equation 1 implies that $s_{ij} = 0$. This results in a division by zero in Equation 9. To circumvent this problem, we do not set the weights $w_{ij}$ equal to one, but we instead define the weights $w_{ij}$ as an increasing function of the similarities $s_{ij}$. More specifically, we define $w_{ij} = s_{ij}$.[6] Equation 9 then becomes

$$\hat{\sigma}(\mathbf{x}_1,\ldots,\mathbf{x}_n) = \sum_{i<j} s_{ij} \left\| \mathbf{x}_i - \mathbf{x}_j \right\|^2 - 2\sum_{i<j} \left\| \mathbf{x}_i - \mathbf{x}_j \right\|. \tag{10}$$

Interestingly, there turns out to be a close relationship between on the one hand the problem of minimizing Equation 10 and on the other hand the problem of minimizing Equation 5 subject to Equation 6. This is stated formally in the following proposition.

[6] Hence, $w_{ij}$ increases linearly with $s_{ij}$. This is the most natural way to define $w_{ij}$. If $w_{ij}$ increases slower than linearly with $s_{ij}$, the division by zero problem remains. If $w_{ij}$ increases faster than linearly with $s_{ij}$, there is no penalty for locating two completely non-similar items close to each other in a map. We further note that $w_{ij} = s_{ij}$ is equivalent with $w_{ij} = 1 / d_{ij}$. This is exactly how weights are chosen in the well-known Sammon mapping variant of MDS (Sammon, 1969).

**Proposition 1.**

(i) If $\mathbf{X} = (\mathbf{x}_1, \ldots, \mathbf{x}_n)$ is a globally optimal solution to the problem of minimizing Equation 10, then there exists a positive real number $c$ such that $c\mathbf{X}$ is a globally optimal solution to the problem of minimizing Equation 5 subject to Equation 6.

(ii) If $\mathbf{X} = (\mathbf{x}_1, \ldots, \mathbf{x}_n)$ is a globally optimal solution to the problem of minimizing Equation 5 subject to Equation 6, then there exists a positive real number $c$ such that $c\mathbf{X}$ is a globally optimal solution to the problem of minimizing Equation 10.

The proof of this proposition is provided in the appendix. It follows from the proposition that, under certain conditions, MDS and VOS are closely related. More specifically, the proposition indicates that VOS can be regarded as a kind of weighted MDS with proximities and weights chosen in a special way.

## *Experimental Comparison*

We now present an experimental comparison of three approaches for constructing bibliometric maps. Two approaches rely on MDS, and the third approach relies on VOS. The two MDS approaches differ from each other in the similarity measure they use. One MDS approach uses a direct similarity measure, namely the association strength defined in Equation 1. The other MDS approach uses an indirect similarity measure, namely the cosine defined in Equation 2. From now on, we refer to the two MDS approaches as the MDS-AS approach and the MDS-COS approach. Like the MDS-AS approach, the VOS approach also uses the association strength similarity measure. Because VOS has been developed to be used specifically in combination with this similarity measure, we do not study the use of VOS in combination with other similarity measures.

Below, we first discuss the data sets that we use in our experimental comparison, and we then discuss the results of the comparison. We also briefly consider the phenomenon of circular maps.

### Data Sets

We use three data sets in our experimental comparison. One data set comprises co-citations of authors in the field of information science, another data set comprises co-citations of journals in the social sciences, and the third data set comprises co-occurrences of keywords in the field of operations research. We refer to the data sets as, respectively, the authors data set, the journals data set, and the keywords data set. All three data sets were obtained from the Web of Science database.

The authors data set was collected as follows. We first delineated the field of information science. To do so, we selected the 36 journals that, based on co-citation data, are most closely related to the *Journal of the American Society for Information Science and Technology* (*JASIST*).[7] These journals and *JASIST* itself constituted our set of information science journals. This set of journals is shown in Table 1. Next, we selected all articles with at least 4 citations (excluding self citations) that were published in our set of information science journals between 1999 and 2008. We then counted for each author the number of selected articles.[8] All authors with at least 3 selected articles were included in the authors data set. There were 405 authors that satisfied this criterion. Finally, we counted the number of co-citations of each pair of authors in the authors data set. The co-citation frequency of two authors takes into account all articles published by the authors in our set of information science journals between 1999 and 2008.

[7] The *Journal of the American Society for Information Science and Technology* and its predecessor, the *Journal of the American Society for Information Science*, were treated as a single journal.

TABLE 1. Set of journals used to delineate the field of information science.

| | |
|---|---|
| ACM Transactions on Information Systems | Knowledge Organization |
| Annual Review of Information Science and Technology | Law Library Journal |
| | Learned Publishing |
| Aslib Proceedings | Library and Information Science Research |
| Bulletin of the Medical Library Association | Library Collections Acquisitions and Technical Services |
| College and Research Libraries | |
| Computers and the Humanities | Library Journal |
| Electronic Library | Library Quarterly |
| Information Processing and Management | Library Resources and Technical Services |
| Information Research-An International Electronic Journal | Library Trends |
| | Libri |
| Information Retrieval | Online |
| Information Technology and Libraries | Online Information Review |
| Interlending and Document Supply | Portal-Libraries and the Academy |
| Journal of Academic Librarianship | Proceedings of the ASIS Annual Meeting |
| Journal of Documentation | Program-Electronic Library and Information Systems |
| Journal of Information Science | |
| Journal of Librarianship and Information Science | Reference and User Services Quarterly |
| Journal of Scholarly Publishing | Research Evaluation |
| Journal of the American Society for Information Science and Technology | Scientometrics |
| | Serials Review |

To collect the journals data set, we first selected all journals in the Web of Science database that belong to at least one social science subject category. We then counted the number of co-citations of each pair of journals. We took into account all citations from articles published between 2004 and 2008 to articles published at most 10 years earlier. Finally, we included in the journals data set all journals with more than 25 co-citations. There were 2079 journals that satisfied this criterion.

The keywords data set has already been used in an earlier paper (Van Eck et al., 2010). The data set includes 831 keywords that were automatically identified in the abstracts (and titles) of 7492 articles published in 15 operations research journals between 2001 and 2006. The co-occurrence frequency of two keywords was obtained by counting the number of abstracts in which the keywords both occur.

## Results

For each of the three data sets that we consider, three maps were constructed, one using the MDS-AS approach, one using the MDS-COS approach, and one using the VOS approach. All maps are two-dimensional. MDS was run using the PROXSCAL program in SPSS. Both MDS approaches used ordinal MDS.[9] 100 random starts of the optimization algorithm were used in all three mapping approaches.[10] For the MDS approaches, stress values calculated using Equation 3 are reported in Table 2. The nine maps that were obtained are available online at www.neesjanvaneck.nl/comparison_mds_vos/, where they can be examined in detail using the VOSviewer software (Van Eck & Waltman, in press). The global structure of each of the maps is shown in Figure 1. In this figure, circles are used to indicate the location of an item. The size of a circle reflects an item's total number of co-occurrences. In order to facilitate the interpretation of the maps, items were clustered using a clustering technique.[11] Colors are used to indicate the cluster to which an item belongs.

[8] Author name disambiguation was performed using an algorithm that we have developed ourselves. A few corrections were made manually. Unlike in some other author co-citation studies, all authors of an article were taken into account, not just the first author.

[9] Ties in the data were kept tied. This is sometimes referred to as the secondary approach to ties (Borg & Groenen, 2005). The secondary approach to ties is the default setting in the PROXSCAL program.

TABLE 2. Stress values calculated using Equation 3 for the MDS-AS and MDS-COS approaches.

| | MDS-AS | MDS-COS |
|---|---|---|
| Authors | 0.12 | 0.04 |
| Journals | 0.14 | 0.05 |
| Keywords | 0.16 | 0.07 |

As can be seen in Figure 1, the MDS-AS, MDS-COS, and VOS approaches produce quite different maps. Although all three approaches succeed to some extent in separating items belonging to different clusters, the global structure of the maps produced by the three approaches is very different. The MDS-AS approach produces maps with the shape of an almost perfect circle. The distribution of items within a circle is more or less uniform, in particular when the number of items is large, as in the case of the journals and keywords data sets. The maps produced by the MDS-COS approach also seem to have a tendency to be somewhat circular, but this effect is much weaker than in the case of the MDS-AS approach. A notable property of the maps produced by the two MDS approaches is that important items (i.e., items with a large number of co-occurrences) tend to be located toward the center of a map. This is especially clear in the case of the authors and keywords data sets. Many relatively unimportant items are scattered throughout the periphery of a map. In the maps produced by the VOS approach, no effects are visible similar to those observed in the case of the two MDS approaches. Hence, the VOS approach does not seem to have a tendency to produce circular maps. It also does not seem to locate important items toward the center of a map. Instead, the VOS approach seems to produce maps in which important and less important items are distributed fairly evenly over the central and peripheral areas.

[10] In the case of the MDS-AS approach, rather stringent convergence criteria were required for the optimization algorithm. Without such criteria, the algorithm was very sensitive to local optima. Due to the stringent convergence criteria, the application of the MDS-AS approach to the journals data set took more than two days of computing time on a standard desktop computer. For comparison, the application of the VOS approach to the same data set took less than ten minutes of computing time.

[11] The clustering technique that was used is similar to the technique discussed in Section 2.3 of a paper by Zhu, Takigawa, Zeng, & Mamitsuka (2009).

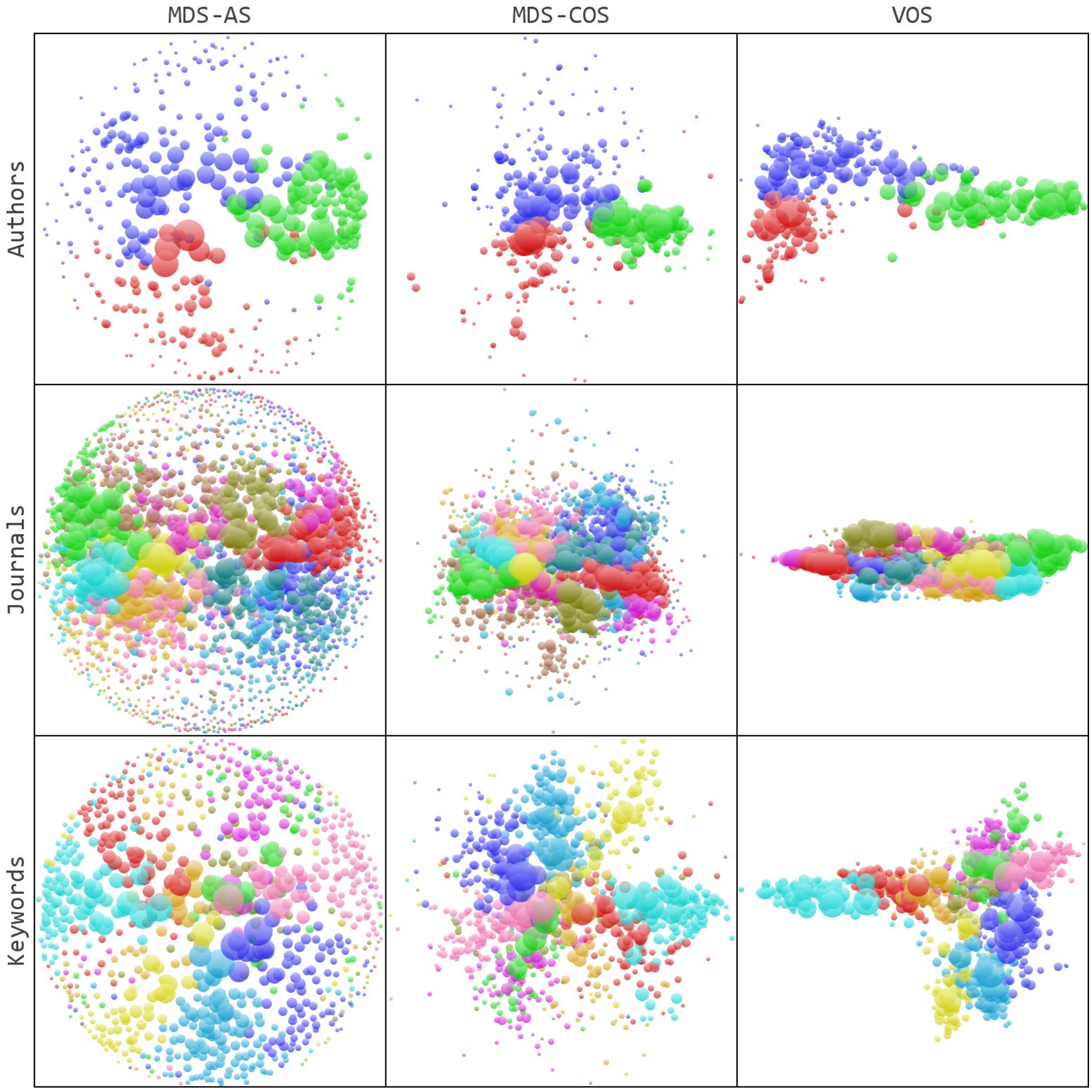

FIG. 1. Global structure of nine maps. Each row corresponds with a data set. Each column corresponds with a mapping approach.

We emphasize that the results shown in Figure 1 are quite robust. The results do not change much when interval MDS is used rather than ordinal MDS. Using MDS combined with direct similarity measures other than the association strength also does not have much effect on the results. Furthermore, the results shown in Figure 1 are relatively independent of the data sets that we use. We investigated numerous other data sets, and this yielded very similar results. However, the almost perfectly circular structure of maps produced by the MDS-AS approach was not observed in the case of data sets with only a relatively small number of items (e.g., less than 100 items). In the bibliometric mapping literature, a clear example of a circular map produced by MDS can be found in a study by Blatt (2009). Blatt used a data set of almost 5000 items. Most bibliometric mapping studies reported in the literature rely on data sets with a much smaller number of items. In such studies, MDS typically does not

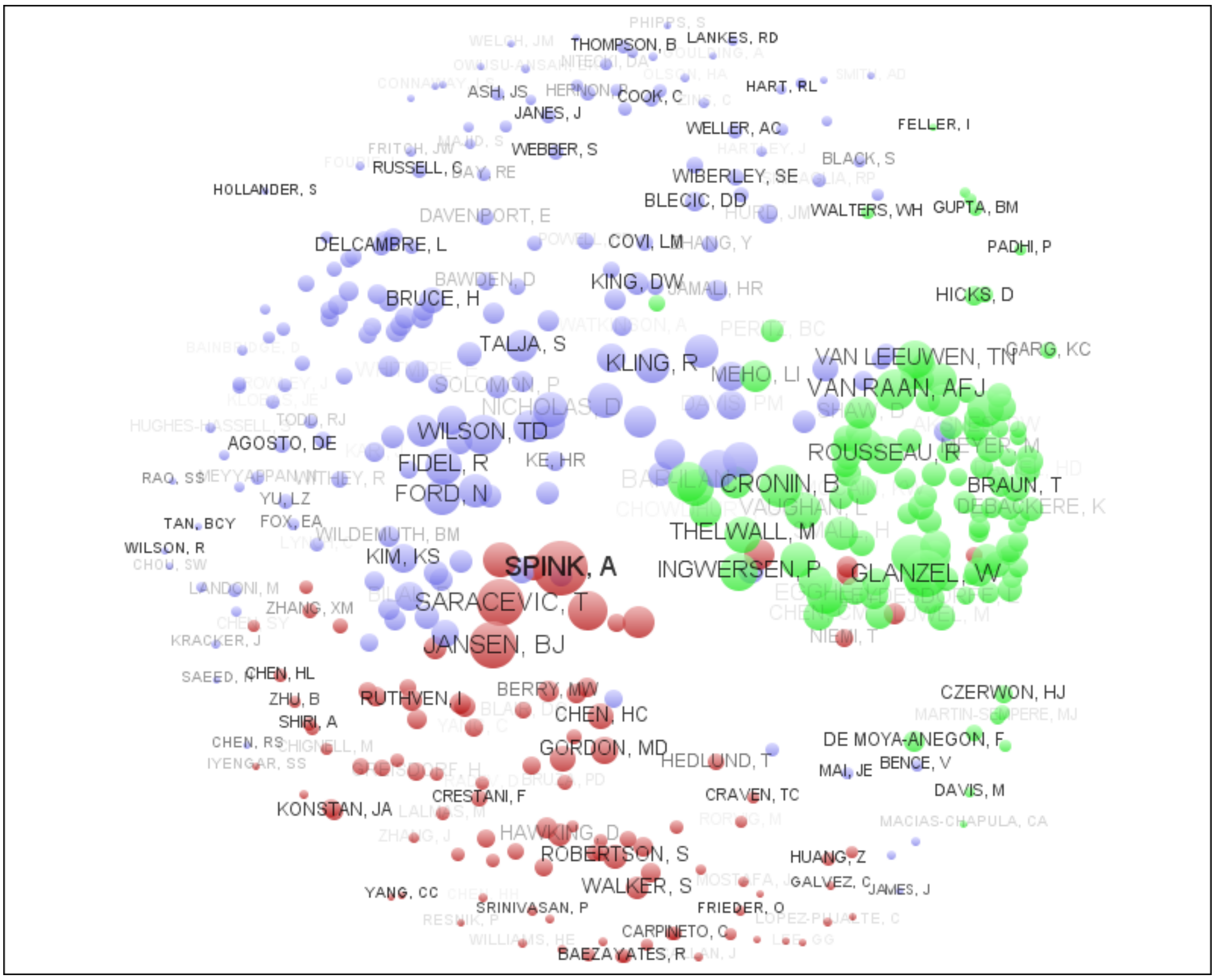

FIG. 2. Map of the authors data set constructed using the MDS-AS approach.

produce circular maps, although a tendency toward a circular structure sometimes seems visible.[12]

We now focus on one data set in more detail. This is the data set of authors in the field of information science. We note that somewhat similar data sets have also been analyzed in a paper by Persson (1994), in a well-known study by White and McCain (1998), and more recently in the work of Zhao and Strotmann (2008a,b,c). Maps of the authors data set constructed using the MDS-AS, MDS-COS, and VOS approaches are shown in Figures 2, 3, and 4, respectively. These are the same maps as the ones shown in the top row of Figure 1.

In various studies of the field of information science (e.g., Åström, 2007; White & McCain, 1998; Zhao & Strotmann, 2008a,b,c), it has been found that the field consists of two quite independent subfields. We adopt the terminology of Åström (2007) and refer to the subfields as information seeking and retrieval (ISR) and informetrics. Comparing the maps in Figures 2, 3, and 4, it can be observed that the separation of the subfields is clearly visible in the VOS map, somewhat less visible in the MDS-COS map, and least visible in the MDS-AS map.[13] In the VOS map, the right part represents the informetrics subfield (e.g., Egghe, Glänzel, and Van Raan) and the left part represents the ISR subfield (e.g., Baeza-Yates, Jansen, Robertson, Spink, Tenopir, and Wilson). There is only a relatively weak connection between the subfields. In the MDS-COS map, the middle right part represents the informetrics subfield and the rest of the map represents the ISR subfield. A striking property of the map is that the ISR subfield is rather scattered, with the most prominent authors (in terms of the number of co-citations) appearing in the center of the map and many somewhat less prominent authors appearing in the periphery. In the MDS-AS map, the middle right part represents the informetrics subfield and the rest of the map represents the ISR subfield. As noted earlier, the map has the shape of an almost perfect circle. The informetrics subfield is partly surrounded by the ISR subfield, with some empty space indicating the separation of the subfields. Prominent authors in the ISR subfield are located toward the center of the map. Less prominent authors tend to be located in the top and bottom parts of the map. This is quite similar to the MDS-COS map.

[12] We note that MDS is not the only mapping technique with a tendency to produce circular maps. See for example Boyack et al. (2005), Heimeriks, Hörlesberger, and Van den Besselaar (2003), Klavans and Boyack (2006), and Noll, Fröhlich, and Schiebel (2002).

[13] In the maps, the green cluster corresponds with the informetrics subfield and the blue and red clusters correspond with the ISR subfield.

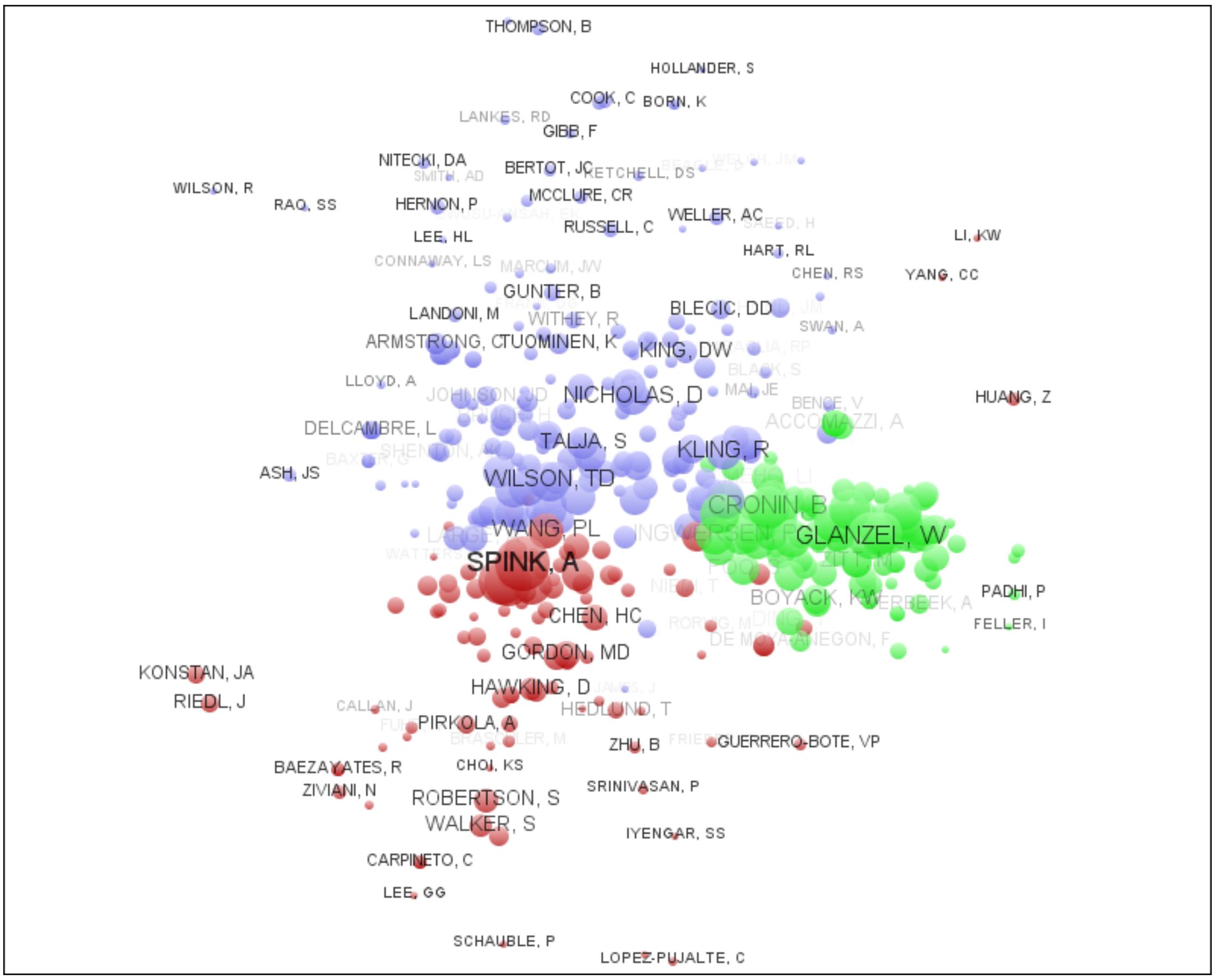

FIG. 3. Map of the authors data set constructed using the MDS-COS approach.

A distinction is sometimes made between “hard” and “soft” ISR research (e.g., Åström, 2007; Persson, 1994; White & McCain, 1998). Hard ISR research is system-oriented and is for example concerned with the development and the experimental evaluation of information retrieval algorithms. Soft ISR research, on the other hand, is user-oriented and studies for example users’ information needs and information behavior. The distinction between hard and soft ISR research is visible in all three maps. In the VOS map, the lower left part represents hard ISR research (e.g., Baeza-Yates and Robertson) and the middle left and upper left parts represent soft ISR research (e.g., Jansen, Spink, Tenopir, and Wilson). In the MDS-COS and MDS-AS maps, the lower part represents hard ISR research and the middle and upper parts represent soft ISR research. As can be seen from all three maps, there is much more soft ISR research than hard ISR research. This is similar to what was found by Åström (2007).

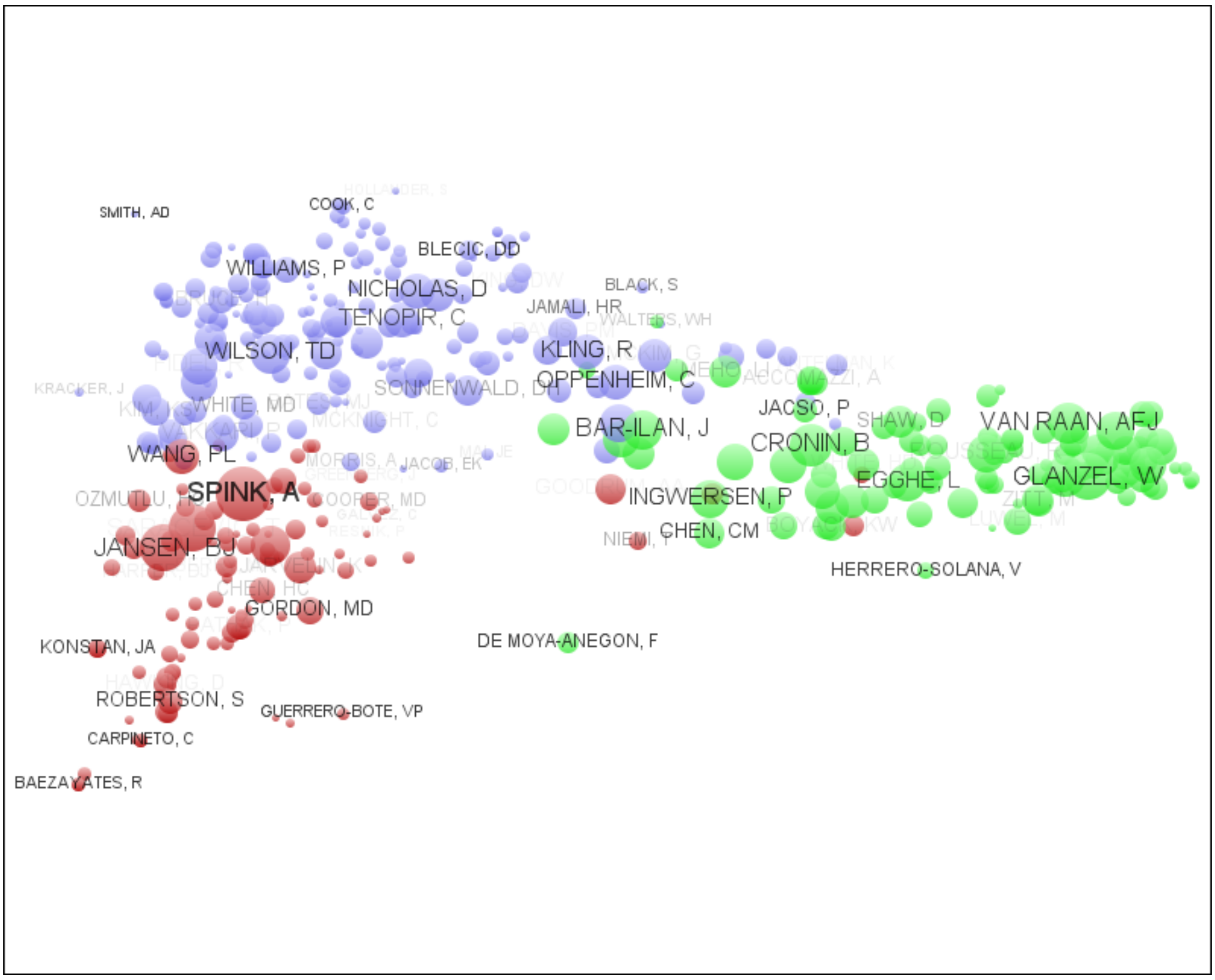

FIG. 4. Map of the authors data set constructed using the VOS approach.

The above comparison of the three maps of the authors data set indicates that the MDS-AS, MDS-COS, and VOS approaches all three succeed reasonably well in locating similar authors close to each other. However, the comparison also makes clear that the MDS-AS and MDS-COS approaches suffer from serious artifacts. Both approaches have a tendency to locate the most prominent authors in the center of a map and less prominent authors in the periphery. Due to this tendency, the separation of subfields becomes more difficult to see. The MDS-AS approach also has a strong tendency to locate authors in a circular structure. This tendency further distorts the way in which a field is represented. Unlike the two MDS approaches, the VOS approach does not seem to suffer from artifacts. That is, the VOS approach does not seem to impose any artificial structure on a map. Our findings based on the maps of the authors data set are confirmed when examining the maps of the journals and keywords data sets. A detailed discussion of the latter maps is beyond the scope of this paper. We note, however, that an examination of these maps indicates the same artifacts of the MDS-AS and MDS-COS approaches as discussed above. The interested reader can verify this at www.neesjanvaneck.nl/comparison_mds_vos/.

The maps in Figures 2 and 3 indicate the consequences of the artifacts from which the MDS-AS and MDS-COS approaches suffer. In these maps, a number of prominent ISR authors (e.g., Spink, Wang, and Wilson) are located equally close or even closer to various informetrics authors than to some of their less prominent ISR colleagues. However, contrary to what the maps seem to suggest, there is in fact very little interaction between the prominent ISR authors and the informetrics authors. The relatively small distance between these two groups of authors therefore does not properly reflect the structure of the field of information science. The small distance is merely a technical artifact, caused by the tendency of the MDS-AS and MDS-COS approaches to locate important items in the center of a map. It follows from this observation that distances in maps constructed using the MDS approaches may not always give an accurate representation of the relatedness of items. Hence, in the case of the MDS approaches, the validity of the interpretation of a distance as an (inverse) measure of relatedness seems questionable. The VOS map in Figure 4 does properly reflect the large separation between the prominent ISR authors and the informetrics authors. In this map, the interpretation of a distance as a measure of relatedness therefore seems valid. We note that the journal and keyword maps available online provide similar examples of the consequences of the MDS artifacts.

## Explanation for Circular Maps

Finally, let us consider the phenomenon of the circular maps produced by the MDS-AS approach in somewhat more detail. Although this phenomenon may seem puzzling at first sight, it actually has a quite straightforward explanation.[14] Co-occurrence data typically consists for a large part of zeros. For example, in the case of the authors, journals, and keywords data sets, respectively 73%, 75%, and 89% of all pairs of items have zero co-occurrences. It follows from Equation 1 that, when two items have a co-occurrence frequency of zero, their association strength equals zero as well. This means that in the MDS-AS approach MDS is typically applied to similarity data that consists largely of zeros. MDS attempts to determine the locations of items in a map in such a way that for each pair of items with a similarity of zero the distance between the items is the same. In the case of similarity data that consists largely of zeros, it is not possible to construct a low-dimensional map with exactly the same distance between each pair of items with a similarity of zero. MDS can only try to approximate such a map as closely as possible. Our experiments indicate that the best possible approximation is a map with an almost perfectly circular structure. This is in fact not a very surprising finding, since it is well known in the MDS literature (Buja, Logan, Reeds, & Shepp, 1994; De Leeuw & Stoop, 1984; see also Borg & Groenen, 2005) that MDS produces perfectly circular maps when all similarities between items are equal. In our experiments, not all similarities between items are equal but only a large proportion. The circular structure of our maps is therefore not perfect but almost perfect.

In our experiments, the VOS approach is applied to the same similarity data as the MDS-AS approach. Hence, the VOS approach is also applied to similarity data that consists for a large part of zeros. This raises the question why, unlike the MDS-AS approach, the VOS approach does not produce circular maps. To answer this question, recall how MDS and VOS are related to each other. As discussed earlier, VOS can be regarded as a kind of weighted MDS with proximities and weights chosen in a special way. More precisely, in the case of VOS, the proximity of two items is set equal to the inverse of the similarity of the items. The weight of two items is set equal to the similarity of the items. From this point of view, one can say that the VOS approach distinguishes itself from the MDS-AS approach in that it does not give equal weight to all pairs of items. The VOS approach gives more weight to more similar pairs of items. It gives little weight to pairs of items with a low similarity. As mentioned above, similarity data is typically dominated by low values, in particular by zeros. These low values cause the MDS-AS approach to produce circular maps. In the case of the VOS approach, however, pairs of items with a low similarity receive little weight and therefore have little effect on a map. Because of this, the VOS approach does not produce circular maps.

[14] For an explanation similar to ours, see Martín-Merino and Muñoz (2004).

## Conclusions

VOS is a new mapping technique that is intended as an alternative to the well-known technique of MDS. We have presented an extensive comparison between the use of MDS and the use of VOS for constructing bibliometric maps. Our analysis has been partly theoretical and partly experimental. In our theoretical analysis, we have studied the mathematical relationship between MDS and VOS. We have shown that VOS can be regarded as a kind of weighted MDS with proximities and weights chosen in a special way. In our experimental analysis, we have compared three approaches for constructing bibliometric maps, two approaches relying on MDS and one approach relying on VOS. We have found that maps constructed using the VOS approach provide a more satisfactory representation of the underlying data set than maps constructed using either of the MDS approaches. The somewhat disappointing performance of the MDS approaches is due to two artifacts from which these approaches suffer. One artifact is the tendency to locate the most important items in the center of a map and less important items in the periphery. The other artifact is the tendency to locate items in a circular structure. Unlike the MDS approaches, the VOS approach does not seem to suffer from artifacts. It is worth emphasizing that our experimental findings are quite robust. We have made the same findings for three fairly different data sets. These data sets differ from each other in size (405, 831, or 2079 items), in type of item (authors, journals, or keywords), and in concept of similarity (co-citation in a reference list or co-occurrence in an abstract).

The interested reader who would like to try out the VOS approach to bibliometric mapping can easily do so using the VOSviewer software (Van Eck & Waltman, in press) that is freely available at www.vosviewer.com. The software offers a graphical user interface that provides easy access to the VOS mapping technique. In addition, the software also comprehensively supports the visualization and interactive examination of bibliometric maps.

## Acknowledgment

We would like to thank Patrick Groenen for his comments on an earlier draft of this paper.

## Appendix

In this appendix, a proof of Proposition 1 is provided. The two parts of the proposition will be proven separately. Both parts will be proven by contradiction.

First consider part (i) of Proposition 1. Let $\mathbf{X} = (\mathbf{x}_1, \ldots, \mathbf{x}_n)$ denote a globally optimal solution to the problem of minimizing Equation 10, and let $\mathbf{Y} = (\mathbf{y}_1, \ldots, \mathbf{y}_n)$ denote a globally optimal solution to the problem of minimizing Equation 5 subject to Equation 6. Let $c$ be given by

$$c = \frac{n(n-1)}{2\sum_{i<j}\left\|\mathbf{x}_i - \mathbf{x}_j\right\|}. \tag{11}$$

Furthermore, define $\mathbf{U} = c\mathbf{X}$ and $\mathbf{V} = \mathbf{Y} / c$. It follows from Equation 11 that $\mathbf{U}$ satisfies the constraint in Equation 6. Assume that $\mathbf{U}$ is not a globally optimal solution to the problem of minimizing Equation 5 subject to Equation 6. This assumption implies that

$$\sum_{i<j} s_{ij}\left\|\mathbf{u}_i - \mathbf{u}_j\right\|^2 > \sum_{i<j} s_{ij}\left\|\mathbf{y}_i - \mathbf{y}_j\right\|^2. \tag{12}$$

It then follows that

$$\sum_{i<j} s_{ij} \left\|\mathbf{x}_i - \mathbf{x}_j\right\|^2 > \sum_{i<j} s_{ij} \left\|\mathbf{v}_i - \mathbf{v}_j\right\|^2 . \tag{13}$$

Extending both the left-hand side and the right-hand side of this inequality with an additional term, where the additional term in the left-hand side equals the additional term in the right-hand side, yields

$$\sum_{i<j} s_{ij} \left\|\mathbf{x}_i - \mathbf{x}_j\right\|^2 - 2\sum_{i<j} \left\|\mathbf{x}_i - \mathbf{x}_j\right\| > \sum_{i<j} s_{ij} \left\|\mathbf{v}_i - \mathbf{v}_j\right\|^2 - 2\sum_{i<j} \left\|\mathbf{v}_i - \mathbf{v}_j\right\| . \tag{14}$$

This inequality implies that **X** is not a globally optimal solution to the problem of minimizing Equation 10. However, this contradicts the way in which **X** was defined. Consequently, the assumption that **U** is not a globally optimal solution to the problem of minimizing Equation 5 subject to Equation 6 must be false. This proves part (i) of Proposition 1.

Now consider part (ii) of Proposition 1. This part will be proven in a similar way as part (i). Let $\mathbf{X} = (\mathbf{x}_1, \ldots, \mathbf{x}_n)$ denote a globally optimal solution to the problem of minimizing Equation 5 subject to Equation 6, and let $\mathbf{Y} = (\mathbf{y}_1, \ldots, \mathbf{y}_n)$ denote a globally optimal solution to the problem of minimizing Equation 10. Let $c$ be given by

$$c = \frac{2\sum_{i<j} \left\|\mathbf{y}_i - \mathbf{y}_j\right\|}{n(n-1)} . \tag{15}$$

Furthermore, define $\mathbf{U} = c\mathbf{X}$ and $\mathbf{V} = \mathbf{Y} / c$. It follows from Equation 15 that **V** satisfies the constraint in Equation 6. Assume that **U** is not a globally optimal solution to the problem of minimizing Equation 10. This assumption implies that

$$\sum_{i<j} s_{ij} \left\|\mathbf{u}_i - \mathbf{u}_j\right\|^2 - 2\sum_{i<j} \left\|\mathbf{u}_i - \mathbf{u}_j\right\| > \sum_{i<j} s_{ij} \left\|\mathbf{y}_i - \mathbf{y}_j\right\|^2 - 2\sum_{i<j} \left\|\mathbf{y}_i - \mathbf{y}_j\right\| . \tag{16}$$

In this inequality, the second term in the left-hand side equals the second term in the right-hand side. The inequality can therefore be simplified to

$$\sum_{i<j} s_{ij} \left\|\mathbf{u}_i - \mathbf{u}_j\right\|^2 > \sum_{i<j} s_{ij} \left\|\mathbf{y}_i - \mathbf{y}_j\right\|^2 . \tag{17}$$

It then follows that

$$\sum_{i<j} s_{ij} \left\|\mathbf{x}_i - \mathbf{x}_j\right\|^2 > \sum_{i<j} s_{ij} \left\|\mathbf{v}_i - \mathbf{v}_j\right\|^2 . \tag{18}$$

This inequality implies that **X** is not a globally optimal solution to the problem of minimizing Equation 5 subject to Equation 6. However, this contradicts the way in which **X** was defined. Consequently, the assumption that **U** is not a globally optimal solution to the problem of minimizing Equation 10 must be false. This proves part (ii) of Proposition 1. The proof of the proposition is now complete.